\documentstyle[aas2pp4]{article}
\begin{document}
\title{NOISE-ENHANCED PARAMETRIC RESONANCE IN PERTURBED GALAXIES}
\author{IOANNIS V. SIDERIS} 
\affil{Department of Physics, Northern Illinois University, DeKalb, IL
60115}
\and
\author{HENRY E. KANDRUP}
\affil{Department of Astronomy, Department of Physics, and Institute for
Fundamental Theory, University of Florida, Gainesville, FL 32611}

\begin{abstract}
This paper describes how parametric resonances associated with a galactic
potential subjected to relatively low amplitude, strictly periodic 
time-dependent perturbations can be impacted by pseudo-random variations 
in the pulsation frequency, modeled as coloured noise. One aim thereby is 
to allow for the effects of a changing oscillation 
frequency as the density distribution associated with a galaxy evolves 
during violent relaxation. Another is to mimic the possible effects of 
internal substructures, satellite galaxies, and or a high density environment. 
The principal conclusion is that allowing for a variable frequency does not
vitiate the effects of parametric resonance; and that, in at least some cases, 
such variations can increase the overall importance of parametric resonance 
associated with systematic pulsations.
\end{abstract}


\section{Introduction and Motivation}
A real galaxy is never a completely isolated time-independent equilibrium. To 
a lesser or greater extent it will always be subjected to a variety of 
perturbations. These include (nearly) regular perturbations, 
reflecting, {\em e.g.,} companion objects and/or (quasi-)normal modes triggered
by close encounters with other galaxies ({\em e.g.,} Vesperini \& Weinberg 
2000) as well as more irregular perturbations, reflecting {\em e.g.,} internal
substructures and/or larger numbers of nearby galaxies in a rich cluster. 
However, recent work ({\em e.g.,} Kandrup, Vass \& Sideris 2003)
has demonstrated that periodic and/or near-periodic 
perturbations can trigger parametric resonances which make a significant 
fraction of the stars execute chaotic orbits, leading to chaotic phase 
mixing and, oftentimes, readjustments in the bulk density distribution.

Most work hitherto on the effects of such parametric resonance has 
assumed a time-dependence
characterised by a single fixed frequency.
However, the frequency may not be exactly constant.
Bulk changes in 
the density distribution during an epoch of violent relaxation will cause
changes in the bulk potential and, hence, the natural pulsation frequency.
And similarly, for a galaxy near equilibrium, even if most of the 
time-dependence is associated with a single normal mode with a well-defined
frequency, one may have to allow for more irregular perturbations reflecting,
{\em e.g.,} nearby galaxies or internal substructures. 

In principle, these
perturbations are deterministic. However, to the extent
that they act in a `near-random' fashion, one might wish to model them as
coloured noise, {\em i.e.,} random `kicks' of finite duration. As emphasised
in Siopis \& Kandrup (2000), any noisy process can be viewed as a superposition
of periodic influences combined with random phases. 

These considerations lead naturally to the issue considered in this paper,
namely how parametric resonance associated with a single well-defined frequency
can be impacted by perturbations inducing random fluctuations in the 
pulsation frequency. Do such fluctuations tend generically 
to enhance the effects of parametric resonance, or do they tend instead to 
weaken its effects? In particular, are there circumstances where even 
small random perturbations can dramatically increase the importance of 
parametric resonance and lead to potentially observable effects? 

One principal conclusion is that a variable 
frequency need not vitiate the effects of parametric resonance; and, that, in
some cases, such variations can make the effects of the resonance more 
pronounced. 
In particular, {\em even for comparatively low amplitude perturbations,
`fuzzing out' the frequency can lead to the displacement of appreciable 
quantities of stars and/or gas to larger radii and, in some cases, to 
their ejection from the galaxy.} This suggests that even comparatively
weak interactions between galaxies ({\em e.g.,} the galactic `harassment'
of Moore, Lake \& Katz [1998]) could contribute
to the material often observed in intergalactic space in rich clusters, an 
effect normally attributed ({\em e.g.,} Bertin 2000) to galaxy-galaxy 
collisions and close encounters. This 
is an effect likely to be missed in $N$-body simulations, where the number of 
`macroparticles' is small compared with the number of stars in a real galaxy.

\section{What Was Done}
As a simple example, one can consider orbits in a galaxy idealised as 
a spherically symmetric Dehnen (1993) model. Allowing for a 
time-dependent perturbation $m(t)$, suppose that, in appropriate units, 
\begin{equation}
V(r,t)=-{m(t)\over (2-{\gamma})}\left[ 1- 
\left( {r\over 1+r} \right)^{2-{\gamma}}\right],
\end{equation}
with $r^{2}=x^{2}+y^{2}+z^{2}$. Attention here will focus 
on ${\gamma}=0$, ${\gamma}=1$, and ${\gamma}=3/2$, corresponding to galaxies 
with no cusp, a moderate cusp, and a steep cusp.

Now suppose further that, in the absence of perturbations, 
$m(t){\;}{\equiv}{\;}1$, and, in a first approximation,
allow for a periodic time-dependent perturbation 
\begin{equation}
m(t)=1+{\delta}m(t)=1+m_{0}\sin {\omega}_{0}t.
\end{equation}
For suitable choices of $m_{0}$ and ${\omega}_{0}$, 
such a perturbation will trigger a parametric resonance which 
can make orbits chaotic and, in some albeit not all such chaotic orbits, 
induce significant changes in energy.

The objective then is to let the frequency to vary in a `random' fashion 
which can be approximated as Gaussian coloured noise with a finite 
autocorrelation time. Assuming that 
${\omega}(t)={\omega}_{0}+{\delta}{\omega}(t)$, with ${\delta}{\omega}$ 
sampling a stationary Ornstein-Uhlenbeck process ({\em e.g.,} Chandrasekhar 
1943, Van Kampen 1981), the random process is characterised 
completely by the first two moments, 
\begin{displaymath}
{\langle}{\delta}{\omega}(t){\rangle} = 0
\end{displaymath}
and
\begin{equation}
{\langle}{\delta}{\omega}(t_{1}){\delta}{\omega}(t_{2}){\rangle} = 
K(|t_{1}-t_{2}|),
\end{equation}
where ${\langle}\;.\; {\rangle}$ denotes a statistical average and the 
autocorrelation function
\begin{equation}
K(|t_{1}-t_{2}|) = {\Delta}^{2}\exp(-|t_{1}-t_{2}|/t_{c})
\end{equation}
Here ${\Delta}$ represents the typical `size' of the random component of
${\omega}$ and $t_{c}$ the time scale over which it changes appreciably.
Alternatively, perhaps more physically, the noise process can be characterised
by $t_{c}$ and a new quantity ${\langle}|{\delta}{\omega}|{\rangle}$, which 
corresponds to the mean value of $|{\delta}{\omega}(t)|$.

The specific focus here was to understand how such random frequency variations
change representative orbits evolved in an otherwise strictly time-periodic 
potential. The first task, therefore, was to explore how initial conditions 
corresponding, in the absence of perturbations, to purely regular, {\em i.e.,} 
nonchaotic, orbits are impacted by an exactly periodic driving.
This involved determining the extent to which the values of
radius $r$ and energy $E$ were affected by the 
perturbation. In the absence of the driving, energy $E$ is of course conserved.
However, in the presence of driving, this is no longer true and, for 
appropriate choices of $m_{0}$ and ${\omega}_{0}$, parametric resonance can 
induce large energy shifts. 

Having determined the effects of such periodic driving, the second, and 
principal, task was to select `interesting' values of unperturbed driving 
frequency ${\omega}_{0}$ and explore how the 
dynamics is changed if
${\omega}$ is perturbed by allowing for random noises with different values of
$m_{0}$, ${\langle}|{\delta}{\omega}|{\rangle}$, and $t_{c}$. 
Most of the experiments assumed $t_{c}=80$. Noting, {\em e.g.,}
that, for the ${\gamma}=1$ Dehnen model, an intermediate energy orbit with 
$E=-0.4$ has 
$t_{D}{\;}{\sim}{\;}10$ (Merrritt \& Fridman 1996), this corresponds to a time 
longer than but still comparable to an orbital time scale. A more systematic 
investigation of variable $t_{c}$ will be presented elsewhere. 

Since ${\delta}{\omega}$ is a random variable, a single integration of an 
initial condition does not suffice. Rather, for each choice of 
initial condition, $m_{0}$, ${\omega}_{0}$, and 
${\langle}|{\delta}{\omega}|{\rangle}$, 10000 different realisations were 
performed with the subsequent analysis focused on extracting statistical 
properties of the resulting orbits. The most important quantities to be 
extracted were the distributions of $r_{max}$ and $E_{fin}$,
the largest value of $r$ 
and final value of $E$ assumed by each of the noisy orbits, which
provide information about the extent to which the noise-enhanced parametric
resonance causes stars to be displaced to larger radii. (The quantity $r_{max}$
would seem a better quantity to track then $r_{fin}$ since, even if the energy
increases monotonically, the radial coordinate will oscillate; and it is
quite possible that, at the end of the integration, $r$ is near a `relative'
perigalacticon.) Attention focused separately on initial conditions 
corresponding to (a) purely radial orbits and (b) orbits with a more isotropic 
distribution of velocities, earlier 
work (Terzi{\'c} \& Kandrup 2003) having shown that non-noisy parametric 
resonance can impact radial and nonradial orbits differently.

This analysis is similar in spirit to recent work on noisy orbits in
charged particle beams (Bohn \& Sideris 2003), where the perturbations 
which one envisions reflect, {\em e.g.,} imperfections in the confining
magnetic field and other details of an external environment. However, the
work described here is different in at least two important respects. 
Firstly, the presence of a confining potential places an upper limit on the 
largest `radius' that, in the absence of noise, a charged particle in the beam 
can achieve, which implies in turn that, for ${\delta}{\omega}=0$, the 
response is a much smoother function of ${\omega}_{0}$ than is the case
for a galaxy. Secondly, and even more importantly, in the context of beam 
dynamics the principal issue of interest is in determining the largest values 
of $r$ and $E$ that any single beam particle can achieve, since the
overriding concern is to prevent resonating particles
from striking the walls of the accelerator. In the context of galactic
dynamics, the actual distribution of maximum radii and energies would seem 
more important. Knowledge of these distributions can, {\em e.g.,} 
facilitate predictions about expected changes in the density distribution 
and/or estimates of the number of stars that might be ejected 
into intergalactic space.

\section{What Was Found}
Consider first orbits subjected to 
strictly periodic driving. In this case, for 
fixed amplitude $m_{0}$ and autocorrelation time $t_{c}$, the response to the 
time-dependence, as probed by $r_{max}$ and $E_{fin}$, can exhibit a 
comparatively sensitive dependence on the frequency ${\omega}_{0}$. Because 
the response reflects a resonant coupling between (harmonics of) the driving 
frequency and the natural frequencies of the orbits, for sufficiently low and 
high ${\omega}_{0}$ the driving has a comparatively minimal effect. In 
particular, the orbits are regular, {\em i.e.,} nonchaotic, as was the case 
for the unperturbed integrable model. For intermediate values, however, there 
are intervals where the orbits become chaotic and the response can be
quite large. 

For smaller amplitudes $m_{0}$, the dependence of quantities like $r_{max}$ 
on ${\omega}_{0}$ tends to be comparatively smooth, successive peaks in 
$r_{max}({\omega}_{0})$, corresponding to strongly chaotic orbits, being
separated by fixed intervals. However, for higher amplitudes the response 
can be more complex, especially for orbits in cuspy potentials. As noted 
elsewhere (Terzi{\' c} \& Kandrup 2003), this is because 
Fourier spectra of cuspy orbits typically exhibit
more structure than spectra of orbits in potentials with a smooth core.

Examples of this behaviour for orbits in pulsed Dehnen potentials with 
${\gamma}=0$ and ${\gamma}=3/2$ are shown in Figure 1, which plots 
$r_{max}({\omega}_{0})$ for initial conditions evolved for a time $t=512$. 
For each potential, results for two initial conditions are exhibited. Both 
correspond to an orbit with initial kinetic energy $T=-E/2$. However, in
one case the initial velocity had components $v_{x}=v_{y}=v_{z}$; in the other 
the velocity was selected to yield a radial orbit. In each case, the initial 
$r=0.25$, this corresponding to a star situated near the center of the galaxy. 
For the cuspless potential, the responses for these two initial conditions are 
quite similar, but for the cuspy potential the radial orbit was often ejected 
to a much larger radius. This is consistent with Terzic \& Kandrup (2003), who
found that, especially for cuspy potentials, radial orbits tended to be 
more susceptible to periodic driving.

Now allow the frequency to vary by including colored noise in
the computations, supposing that the unperturbed 
${\omega}_{0}$ is in one of the resonant regions.
At least for `reasonable' choices of parameter values, {\em i.e.,} 
${\langle}|{\delta}{\omega}|{\rangle}$ smaller than the width of the 
resonance, allowing the pulsation frequency to 
vary results on the average in 
orbits being displaced to larger radii. However, when 
${\langle}|{\delta}{\omega}|{\rangle}$ is very small, 
such variations have a comparatively minimal effect.
If ${\langle}|{\delta}{\omega}|{\rangle}/{\omega}_{0}$ is very 
small, say $<10^{-4}$ or so, as a practical matter the frequency is `nearly 
constant' in the sense that the variations typically have no appreciable 
impact on orbital structure, so that nothing really changes. Alternatively,
when ${\langle}|{\delta}{\omega}|{\rangle}$ becomes 
too large these variations can 
suppress the transport of stars to larger radii and less negative energies. 
If the variation is too large, the effective frequency with which the orbit is 
perturbed can
be shifted out of the `interesting' resonant range, at which point the driving
no longer has a significant effect. {\em The important point is that 
finite-sized variations that wiggle the frequency within the resonant range 
tend generically to increase the mean $r_{max}$.}

Examples of this behaviour are exhibited in Figure 2, which focuses on the
same four initial conditions used to construct Figure 1, pulsed with frequency
${\omega}_{0}=0.6$ but allowing also for variations with variable 
${\langle}|{\delta}{\omega}|{\rangle}$. In each case the data derive from
10000 noisy integrations, each proceeding for a time $t=512$. The
solid curve represents the largest radial excursion experienced by any of the
10000 orbits. The dotted curve represents the mean value of $r_{max}$
averaged over the different orbits. For very small and large values of 
${\langle}|{\delta}{\omega}|{\rangle}$, the frequency variations 
have only a comparatively minimal effect but, for 
${\langle}|{\delta}{\omega}|{\rangle}{\;}{\sim}{\;}10^{-3}-10^{-1}$,
the mean effect of this irregular time-dependence is substantially larger than 
for a purely periodic time-dependence. The range of values of 
${\langle}|{\delta}{\omega}|{\rangle}$ resulting in a significant effect is
comparable for radial and nonradial orbits. However, especially for the cuspy 
potential, within this range of values a varying frequency tends to have a 
larger effect on radial orbits. 

As is illustrated in Figure 3, a varying frequency can also increase 
the largest $E_{fin}$. However, the mean value of $E_{fin}$ does {\em not} tend
to grow significantly.

Figure 2 demonstrates that the average effect of a variable frequency is to 
drive orbits to somewhat larger radii. However, it is also evident that 
some orbits can be expelled to {\em very} large values of $r$: If the
orbit and the noise are appropriately `tuned', the resonant coupling can have
an enormous effect. For example, a radial initial condition in the 
${\gamma}=3/2$ Dehnen potential which, in the absence of a frequency 
variations, is restricted
to radii $r<1.2$ can, in the presence of such variations, be ejected to
$r>15$! The obvious question, however, is how rare such events actually are.
To address this issue, and to better understand the variations that
a nonzero ${\langle}|{\delta}{\omega}|{\rangle}$ can trigger, it is useful
to examine $f(r_{max})$, the distribution of values of $r_{max}$ for the 
10000 different noisy realisations.

Several examples of such distributions, generated for orbits in
a ${\gamma}=0$ Dehnen potential with ${\omega}_{0}=0.6$ and $m_{0}=0.05$, are 
presented in Figure 4.
Perhaps the most obvious point here is that these distributions can be quite
complex. In general they are asymmetric, and they often exhibit multiple
peaks, suggestive of a superposition of seemingly distinct populations.
For the lowest amplitude variations exhibited in that Figure, 
${\langle}|{\delta}{\omega}|{\rangle}{\;}{\approx}{\;}0.0002$, 
the mean values of $r_{max}$ for the ensembles are very close to the $r_{max}$ 
computed in the absence of any frequency variations, but the distributions
$f(r_{max})$ still manifests significant structure. In particular, the 
distribution for the radial initial condition exhibits two peaks, the existence
of which is of definite statistical significance. 
For the intermediate value 
${\langle}|{\delta}{\omega}|{\rangle}{\;}{\approx}{\;}0.06$, the shape of
the distribution $f(r_{max})$ is relatively similar to the distributions in
the upper panels. However, in this case most of the orbits have
managed to reach larger radii; only a few are restricted to
values smaller than the $r_{max}$ associated with the driftless orbit.
For the largest amplitude,
${\langle}|{\delta}{\omega}|{\rangle}{\;}{\approx}{\;}0.25$, even though 
the mean change in $r_{max}$ is relatively small, the distribution is quite
broad; and it is clear that a majority of the orbits are actually restricted
to smaller radii than was the driftless orbit.

These effects may perhaps be better understood
by focusing on changes in energy rather than radius. This is illustrated 
in Figure 5, which exhibits $f(E_{fin})$, the distribution of final energies,
computed for the same orbits used to generate Figure 4. 
Even visually, it is evident that, 
for the medium amplitude case, where the effect of the frequency
variation is largest, 
$f(E_{fin})$ is reasonably well fit by a Gaussian with a mean
equal to the $E_{fin}$ associated with the zero variation orbit. It 
thus appears that, at least as viewed in energy space, the 
variations
result in strictly random changes in energy superimposed on the energy shift 
that arises in the absence of the variation. 

For the highest amplitude, the distributions are again Gaussian although
the mean value is substantially smaller than the $E_{fin}$ for 
the driftless orbit. This decrease can likely be attributed to the fact that,
in this case, ${\langle}|{\delta}{\omega}|{\rangle}$ is sufficiently large
that, in many cases, the effective frequency acting on the orbit is outside
the resonant region, thus yielding a weaker response overall. 
(As is evident from Figure 1 a and b, the width of the resonance near 
${\omega}=0.6$ is 
${\la}{\;}0.3$.)

For the lowest amplitude case, the situation is very different. For the
isotropic initial condition, $f(E_{fin})$ has an extended low energy tail; for 
the radial case, it is bimodal, indicative of two distinct 
populations. Given that the initial $E=-0.24$, it is
evident that the total time-dependence (almost) never extracts energy from 
the orbit.
However, the lower energy portion of the distribution corresponds to orbits 
which experience only minimal changes. 

The key to understanding this behaviour is the fact that
chaotic orbits in a strictly periodically pulsed Dehnen potential divide
empirically into two `types', namely `sticky' chaotic orbits which exhibit
little if any systematic changes in energy and `wildly' chaotic orbits which
do exhibit systematic changes (Terzi{\'c} \& Kandrup 2003). Analysis of 
Fourier spectra indicates that the `sticky' orbits are `locked'
to a harmonic of the driving frequency, which restricts the extent to which
they can drift in energy, whereas the `wildly' chaotic are unconstrained.
However, this distinction is not absolute. For example, an orbit that starts 
as `sticky' will eventually become unstuck and start drifting through energy
space in much the same sense that, in a time-independent Hamiltonian system,
a chaotic orbit originally trapped near a regular island can diffuse through
cantori to move into the chaotic sea. This analogy with time-independent
Hamiltonian systems suggests in turn that, just as noise can facilitate
diffusion through cantori, corresponding to transitions between `sticky' and
`wildly' chaotic behaviour (Pogorelov \& Kandrup 1999), the introduction of
weak noise into the driving frequency could facilitate transitions between
orbits which do, and do not, exhibit systematic energy drifts.

If the frequency shifts are large, it might seem unlikely for  `stickiness' 
to persist for long time intervals since this would require the orbital
frequencies to vary in just the right way as to remain locked to the driving.
If, however, the frequency shifts are small, an orbit that started as sticky
might be expected to remain sticky for a relatively long time; and similarly,
if the orbit starts as wildly chaotic, there is the possibility that it will
be `kicked' into a frequency `lock' and remain there for quite a while.

That such behaviour is possible is illustrated in Figure 6, which exhibits
$E(t)$ and $r(t)$ for an orbit evolved in a pulsed ${\gamma}=1$ Dehnen
potential with ${\omega}_{0}=0.6$, $m_{0}=0.01$, and 
${\langle}|{\delta}{\omega}|{\rangle}=0.003$ for a total time $t=2048$. 
Viewed over times $t<500$ or so, the energy does not appear to exhibit any
systematic changes and the radial motion appears comparatively regular, even
though an examination of the Fourier spectra suggests 
that the orbit is not regular. It is, however, evident that,
at a slightly later time, the energy begins to evolve in a more irregular
fashion and that, for $t>1100$ or so, the energy increases systematically.

In any event, the behaviour exhibited in Figure 5 for the medium amplitude 
case becomes progressively more common for larger amplitudes and cuspier 
potentials, where the resonances become broader and `sticky' chaotic orbits
become less common. 

\section{Conclusions and Interpretations}
Periodic driving tends generically to pump energy into the stars in a galaxy,
thus displacing them towards higher energies and larger radii, an effect 
resulting ({\em e.g.,} Kandrup, Vass \& Sideris 2003, Terzi{\'c} \& Kandrup 
2003) from a resonant coupling between the driving frequency and the 
frequencies of the orbits, which also make the orbits chaotic. Even relatively 
weak driving with fractional amplitude ${\la}{\;}0.05$ can have significant 
effects within a few tens of dynamical times; the effects of larger amplitude 
can be large enough to account for violent relaxation (at least in principle).
Given, however, that this shuffling in energies involves a resonance, one
might wonder to what degree this behaviour persists if the driving frequency
is allowed to vary. 

One principal conclusion of this paper is that such variations do not vitiate
the effects of such resonant couplings. Indeed, allowing for modest variations
in frequency tends on average to increase the maximum radii to which orbits
are displaced. However, this does not imply that, on the average, such
variations result in more energy being pumped into the orbits. To the extent
that the orbits have all become `wildly' chaotic, {\em i.e.,} that they have
one or more positive Lyapunov exponents and that they are not `locked' to the
(near-constant) driving frequency, allowing for different random frequency
variations leads to a Gaussian distribution of maximum energies $E_{fin}$ 
centered
about the maximum energy attained by an orbit subjected to strictly periodic
driving. The systematic increase in the average $r_{max}$ arises because
realistic (near-)equilibria have phase space distributions which are
monotonically decreasing functions of energy (${\partial}f/{\partial}E<0$),
so that a symmetric spread in energies occasions an increase in the number
of larger radii stars at the expense of stars at smaller radii.

More importantly, a random frequency can have a {\em very} large impact on at 
least a small number of stars. In particular, it is
statistically probable that a few orbits will experience a noisy variable
frequency which is `just right' to give it a very large energy, thus displacing
it to very large radii. (Alternatively, one might expect analogously that,
for a fixed noisy variable frequency, there will be a small number of initial
conditions resulting in `just right' chaotic orbits which will be impacted
significantly.) In particular, the computations described above show that,
in some cases, even a perturbation of fractional amplitude as weak as 
$m_{0}{\;}{\sim}{\;}0.05$ can cause a star to reach a radius an order of 
magnitude larger than it would have reached if driven with a fixed frequency.

This is an effect likely to be missed in numerical simulations of galaxy
evolution. As illustrated in Figure 7 for $r_{max}$, the distributions of 
values of $r_{max}$ 
and $E_{fin}$ decay exponentially for large values. This implies, {\em e.g.,}
that the largest value of $r_{max}$ computed for a collection of $N$ orbits 
should increase logarithmically in $N$, an effect illustrated in Figure 8. 
Given, however, that few simulations of galaxies incorporate more than 
${\sim}{\;}10^{7}$ macroparticles, it would seem highly probable that such 
simulations could miss rare `events' that arise at least occasionally in 
systems with $>10^{11}$ stars.

The computations described here suggest strongly that subjecting a galaxy
to time-dependent perturbations which can be idealised as nearly periodic
oscillations with variable frequency can have significant, potentially
observable, effects. However, there remain a large number of lacuna, some
of which are currently under investigation (Kandrup, Sideris \& Bohn 2003).
Most obvious, perhaps, is the question of how the results derived here 
depend on the autocorrelation time $t_{c}$. There are indications (Kandrup,
Vass \& Sideris 2003) that the precise value of $t_{c}$ may not be all that 
important, but this remains to be checked. 

Also important is the issue of how coloured noise can facilitate transitions
between `sticky' and `wildly' chaotic orbits. Earlier work (Terzic \& Kandrup 
2003) has shown that chaotic orbits subjected to purely periodic driving 
divide into two `types', separated by `entropy barriers', namely 
`sticky' chaotic orbits which remain in the same phase space region without 
exhibiting systematic secular variations in energy and `wildly' chaotic orbits 
where changes in energy typically grow diffusively. The obvious question, then,
is to what extent the introduction of noise might accelerate transitions 
between orbit types. One knows, for example, that, in a time-independent
Hamiltonian system, noise can facilitate diffusion through cantori or along the
Arnold web (Lichtenberg \& Wood 1989, Pogorelov \& Kandrup 1999).

Practical applications also require that the results described here be
generalised to allow for unperturbed potentials which are nonspherical,
especially those admitting chaotic orbits. Earlier work (Kandrup, Sideris,
Terzi{\'c} \& Bohn 2003) has shown that time-independent potentials that 
already admit 
chaos can be significantly more susceptible to time-periodic perturbations,
especially those characterised by relatively low frequencies (as might be
appropriate for modeling an external environment); and it would not seem
unreasonable to conjecture that frequency variations might also prove more
important in such settings. 

Most importantly, however, it would seem imperative to consider the 
statistical effects of a noisy frequency on ensembles of initial conditions 
that constitute fair phase space samplings, so as to facilitate detailed
predictions 
about the macroscopic properties of `real' galaxies. In the first instance 
this could be done by considering {\em e.g.,} microcanonical samplings of 
constant energy hypersurfaces in fixed potentials. Ultimately, however,
this problem can -- and will -- be addressed in the context of fully 
self-consistent simulations.
\vskip .2in
Thanks to Court Bohn for useful discussions.
HEK was supported in part by NSF AST-0307351. IVS was 
supported in part by Department of Education grant G1A62056.
\vskip .5in
\centerline{REFERENCES}
\par\noindent
Bertin, G. 2000, Dynamics of Galaxies, Cambridge University Press, Cambridge
\par\noindent
Bohn, C.~L., Sideris, I.~V. 2003, Phys. Rev. Lett., submitted
\par\noindent
Chandrasekhar, S. 1943, Rev.~Mod.~Phys. 15, 1
\par\noindent
Dehnen, W. 1993, MNRAS 265, 250
\par\noindent
Kandrup, H.~E., Sideris, I.~V., Bohn, C.~L., 2003, in preparation
\par\noindent
Kandrup, H.~E., Sideris, I.~V., Terzi{\'c}, Bohn, C.~L. 2003, ApJ, in press
\par\noindent
Kandrup, H.~E., Vass, I.~M., Sideris, I.~V. 2003, MNRAS 341, 927
\par\noindent
Lichtenberg, A.~J, Wood, B.~P. 1989, Phys. Rev. Lett. 62, 2213
\par\noindent
Merritt, D., Fridman, T. 1996, ApJ 460, 136
\par\noindent
Moore, B., Lake, G., Katz, N. 1998, ApJ 495, 139
\par\noindent
Pogorelov, I. V., Kandrup, H. E., 1999, Phys. Rev. E 60, 1567
\par\noindent
Siopis, C., Kandrup, H.~E. 2000, MNRAS 319, 43
\par\noindent
Terzi{\'c}, B., Kandrup, H.~E. 2003, MNRAS, submitted
\par\noindent
Van Kampen, N.~G. 1981, Stochastic Processes in Chemistry and
Physics, North Holland, Amsterdam
\par\noindent
Vesperini, E., Weinberg, M.~D. 2000, ApJ 534, 598.

\begin{figure}
\plotone{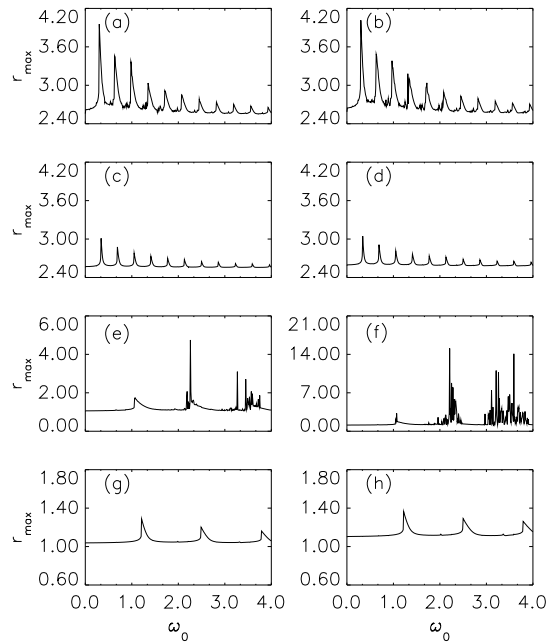}
\caption{(a) $r_{max}$, the maximum value assumed by the radial coordinate 
$r$ for orbits generated from an isotropic initial condition with $E=-0.24$, 
evolved deterministically in a pulsed ${\gamma}=0$ Dehnen potential with 
$m_{0}=0.05$ and variable frequency ${\omega}_{0}$. (b) The same for a radial 
initial condition with $E=-0.24$. (c) The same isotropic initial 
condition evolved with $m_{0}=0.01$. (d) The same radial initial condition
evolved with $m=0.01$. In each case frequency wassampled at intervals
${\delta}{\omega}_{0}=0.01$. (e) - (h)
The same as the preceding, now allowing for initial conditions
with $E=-0.55$ evolved in a ${\gamma}=3/2$ Dehnen potential.}
\end{figure}

\begin{figure}
\plotone{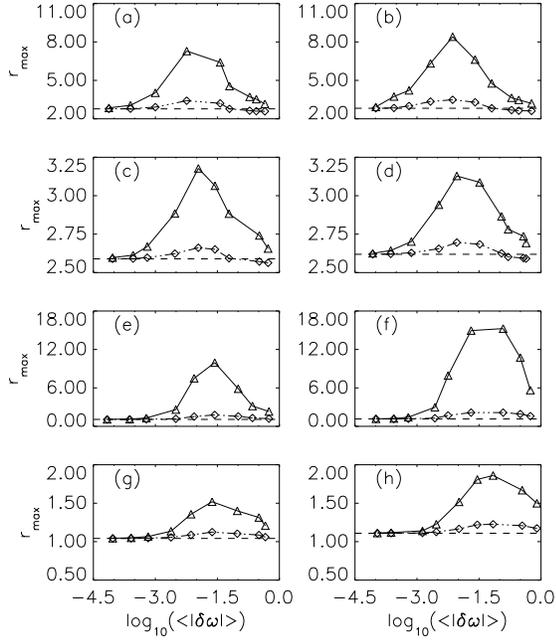}
\caption{(a) The largest value (solid) and the mean value (dotted) of the 
maximum radial coordinate $r_{max}$ for noisy orbits generated from an 
isotropic initial condition with $E=-0.24$, evolved in a ${\gamma}=0$ Dehnen 
potential with ${\omega}_{0}=0.6$, $m_{0}=0.05$, and $t_{c}=80$. 
(b) The same for a radial initial condition with $E=-0.24$. (c) The same 
isotropic initial 
condition evolved with $m_{0}=0.01$. (d) The same radial initial condition
evolved with $m=0.01$. The dashed line corresponds to a deterministic orbit
with ${\delta}{\omega}{\;}{\equiv}{\;}0$. (e) - (h)
The same as the preceding, now allowing for initial conditions
with $E=-0.55$ evolved in a ${\gamma}=3/2$ Dehnen potential.}
\end{figure}

\begin{figure}
\plotone{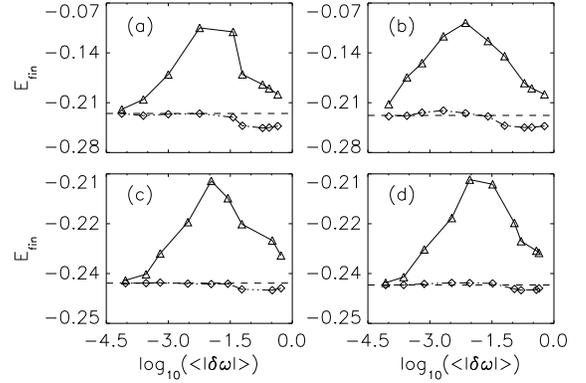}
\caption{The analogue of Figure 2 (a) - (d), now exhibiting the maximum
energy $E_{fin}$ rather than $r_{max}$.}
\end{figure}

\begin{figure}
\plotone{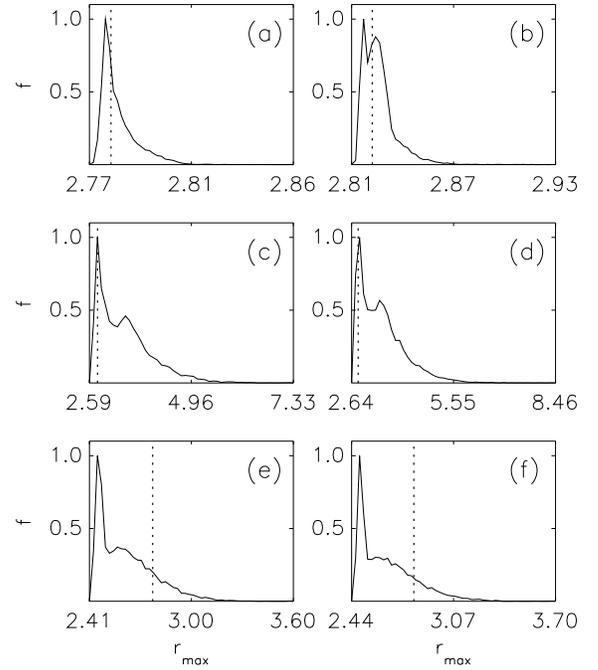}
\caption{The distribution of maximal radial excursions, $r_{max}$ for the
two initial conditions of Fig.~1 evolved in the ${\gamma}=0$ Dehnen potential
with ${\omega}_{0}=0.6$ and $m_{0}=0.05$. The left panels are for the isotropic
initial condition with (top to bottom) 
${\langle}|{\delta}{\omega}|{\rangle}{\;}{\approx}{\;}0.0002$, $0.006$, and 
$0.25$. The right panels are for the radial initial condition with the same 
values of ${\langle}|{\delta}{\omega}|{\rangle}$. The dotted vertical line 
exhibits $r_{max}$ for the same initial condition subjected to purely periodic 
driving.}
\end{figure}

\begin{figure}
\plotone{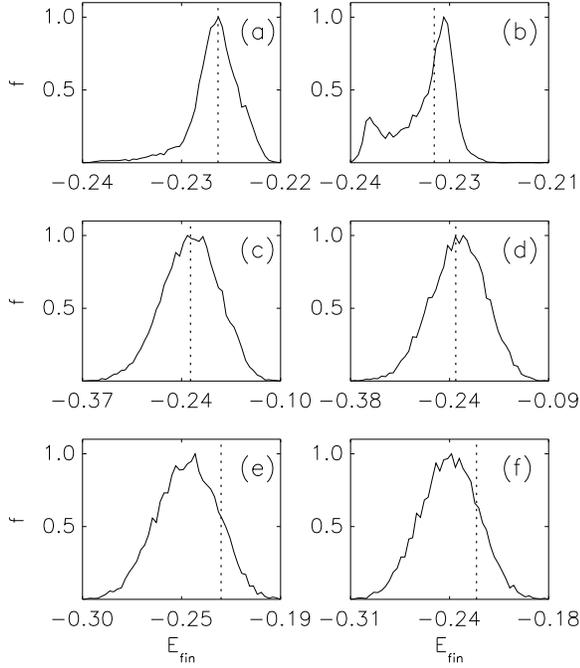}
\caption{The distribution of final energies $E_{fin}$ for the same data
analysed in the preceding Figure. The dotted vertical line represents 
$E_{fin}$ for the same initial condition subjected to purely periodic
driving.}
\end{figure}

\begin{figure}
\plotone{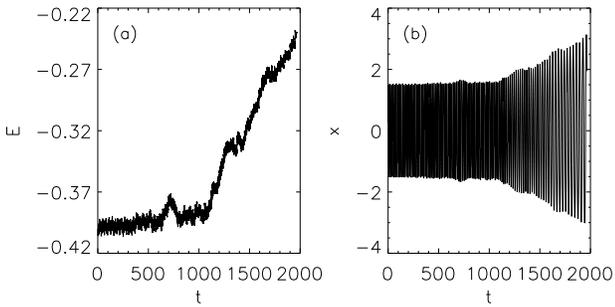}
\caption{(a) The energy $E(t)$ of a radial orbit with $E(0)=-0.4$ and 
$y{\;}{\equiv}{\;}z{\;}{\equiv}{\;}0$ evolved in a pulsed ${\gamma}=1$ 
Dehnen potential with ${\omega}_{0}=0.6$, $m_{0}=0.01$, and 
${\langle}|{\delta}{\omega}|{\rangle}=0.003.$ (b) The coordinate $x(t)$
for the same orbit.}
\end{figure}

\begin{figure}
\plotone{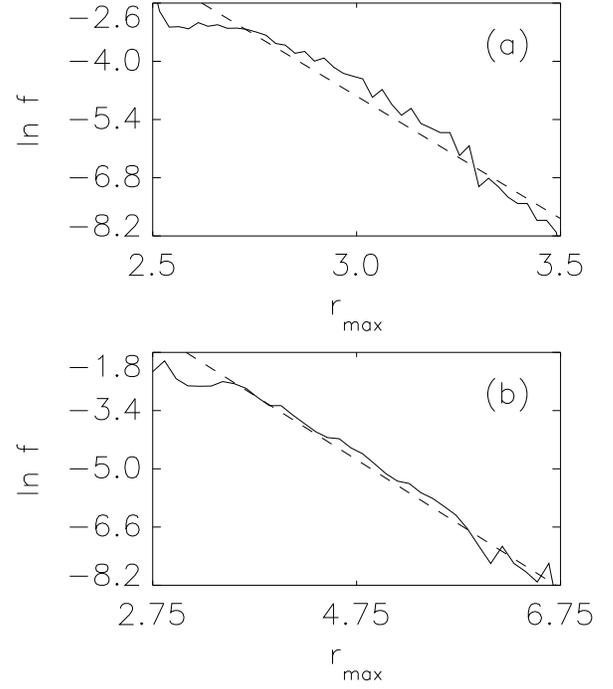}
\caption{(a) Exponential fall-off in the distribution $f(r_{max})$ for an
isotropic initial condition evolved in a pulsed ${\gamma}=0$ Dehnen potential
with ${\omega}=0.6$, $m_{0}=0.05$, and $t_{c}=80$, and 
${\langle}|{\delta}{\omega}|{\rangle}=0.2$. (b) The same for 
${\langle}|{\delta}{\omega}|{\rangle}=0.007$.}
\end{figure}

\begin{figure}
\plotone{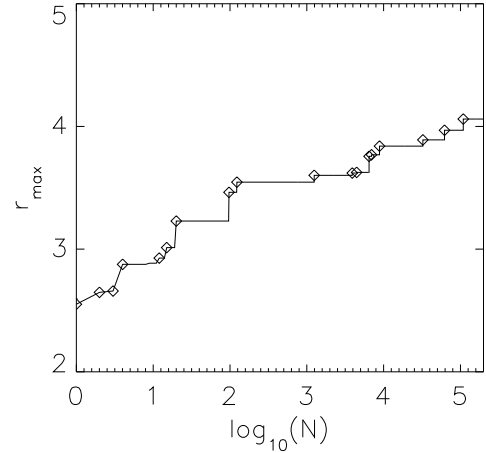}
\caption{The largest value of $r_{max}$ for a radial initial condition 
evolved in a pulsed ${\gamma}=0$ Dehnen potential with ${\omega}=2.8$, $m=0.05$,
$t_{c}=80$, and ${\langle}|{\delta}{\omega}|{\rangle}=0.08$, plotted as a 
function of the number $N$ of noisy realisations.}
\end{figure}

\vfill\eject
\end{document}